# Dynamic Transfer of Chiral Edge States in Topological Type-II Hyperbolic Lattices

Jingming Chen[1], Linyun Yang[2,*], Zhen Gao[1,*]

[1]State Key Laboratory of Optical Fiber and Cable Manufacture Technology, Department of Electronic and Electrical Engineering, Guangdong Key Laboratory of Integrated Optoelectronics Intellisense, Southern University of Science and Technology, Shenzhen 518055, China.
[2]College of Aerospace Engineering, Chongqing University, Chongqing, 400030, China.

*Corresponding author. Email: yanglinyun@cqu.edu.cn (L.Y.Y.); gaoz@sustech.edu.cn (Z.G.)

**Abstract**
The discovery of hyperbolic lattice, a discretized regularization of non-Euclidean space with constant negative curvature, has provided an unprecedented platform to extend topological phases of matter from Euclidean to non-Euclidean spaces. To date, however, all previous hyperbolic topological states are limited to conventional type-I hyperbolic lattice with a single edge, leaving the dynamic transfer of hyperbolic topological states between different edges completely unresolved. Here, by extending the hyperbolic topological physics from the conventional type-I hyperbolic lattices to the newfangled type-II hyperbolic lattices, we report the type-II hyperbolic Chern insulator featuring outer and inner chiral edge states and demonstrate their dynamic transfer across the bulk to the opposite edge via two distinct mechanisms: anti-parity-time phase transition and Landau-Zener single-band pumping. Our work lays the foundation for further exploring the dynamic evolution of hyperbolic topological effects, with the final goal of inspiring applications leveraging dynamic manipulations of the hyperbolic topological states.

**Introduction**

Exploring and discovering novel topological phases of matter has emerged as the most pivotal and fascinating research area in condensed-matter physics [1,2], photonics [3,4], acoustics [5-7], mechanics [7,8], electric circuits [9-11], and even thermal diffusions [12,13]. To date, most of the nontrivial topological states have been established in the Euclidean spaces with zero curvature. Recently, the discovery of hyperbolic lattices, which are the regular tessellations in non-Euclidean spaces with constant negative curvature, in circuit quantum electrodynamics [14] and electric circuits [15] has opened up new avenues to extend topological physics from the Euclidean to non-Euclidean spaces and simulated numerous research advances in hyperbolic topological physics with novel topological states, including the hyperbolic quantum spin Hall effect [16], hyperbolic Haldane [17-19] and Kane Mele models [18], hyperbolic Hofstadter butterfly [20], hyperbolic band topology with second Chern number [21], hyperbolic graphene [22], hyperbolic photonic topological insulators [23], anomalous and Chern edge states in hyperbolic networks [24], hyperbolic non-abelian semimetal [25], higher-order topological hyperbolic lattices [17,26,27] and so on.

However, so far, all previous studies on hyperbolic topological physics have been limited to conventional type-I hyperbolic lattices and only focus on the hyperbolic topological states associated with a single edge [16-27]. In this context, a significant class of topological processes, namely the dynamic transfer of topological states between different edges, such as Thouless pumping [28,29], Laughlin pumping [30-33] and Landau-Zener (LZ) transition [34-37], cannot be achieved in the existing framework of type-I hyperbolic lattices. These dynamic topological effects are crucial for advancing hyperbolic topological physics in that they not only provide a powerful approach to manipulate hyperbolic topological states but also offer a crucial method for characterizing hyperbolic topological invariants and exploring higher-dimensional hyperbolic topological physics [38,39]. Fortunately, the recently discovered type-II hyperbolic lattices [40,41] feature inner and outer edges, thereby offering a natural platform to fill this long-standing gap in hyperbolic topological physics.

In this work, by mapping the celebrated Qi-Wu-Zhang (QWZ) model onto a type-II hyperbolic lattice, we report a type-II hyperbolic Chern insulator (HCI) whose inner and outer edges support counterpropagating chiral edge states (CESs). More significantly, we demonstrate that these CESs can be dynamically transferred between the outer and inner edges of a type-II HCI by leveraging the power-flow conversion at an anti-parity-time symmetric (APT-S) exceptional point (EP) [42] or adiabatic evolution along a single band in LZ model [36-37], which cannot be achieved in type-I HCIs with only a single edge. Our findings not only lay the foundation for future studies on the dynamic transfer of hyperbolic topological states but also inspire potential applications that harness non-Hermitian phase transitions and adiabatic evolution as new degrees of freedom to manipulate hyperbolic topological states, thereby expanding the research scope of hyperbolic topological physics.

**Results**
**Type-I and type-II hyperbolic lattices**

We start with classifying hyperbolic planes with constant negative curvatures in three-dimensional (3D) Minkowski spacetime and their projection onto planar hyperbolic models. In general, hyperbolic planes can be fundamentally categorized as two-sheet (Fig. 1a) and one-sheet (Fig. 1d) hyperboloids (see Supplementary Note 1). Under stereographic projection, these two types of hyperboloids can be further mapped as planar "Poincaré disk" (Fig. 1b) and "Poincaré ring" (Fig. 1e) models, respectively (see Supplementary Note 2), while preserving their local constant negative curvatures and global topologies. More intriguingly, by applying regular polygon tessellation that discretizes a flat plane into a crystal lattice, these two types of Poincaré model can be further discretized as type-I (Fig. 1c) and type-II hyperbolic lattices (Fig. 1f), respectively.

In contrast to the type-I hyperbolic lattice whose geometric degrees of freedom can be fully characterized by a Schläfli symbol $\{p, q\}$, an extended Schläfli symbol $\{r_h, p, q\}$ has to be adopted to label the unique geometric topology of the type-II hyperbolic lattice, where $p$ and $q$ represent $q$ copies of $p$-sided polygons meeting at each vertex, $r_h = e^{-2\pi/kP}$ represents the characteristic radius (radius of the yellow circle in Figs. 1e-1f), $P$ is a geometry constant (see Supplementary Note 3) and $k$ is a positive integer representing the rotation symmetry of the type-II hyperbolic lattice (in this work we fix $k$ as even numbers unless otherwise specified). The inner radius $r_{in}$ can be determined by the characteristic radius $r_h$ via the relation $r_{in} = r_h^2$. Here, we adopt a finite type-II hyperbolic $\{0.365, 8, 3\}$ lattice ($k = 4, P = 1.559$) with a total of 1240 sites.

**Type-II hyperbolic Chern insulators**

To construct a type-II HCI, as illustrated in Fig. 2a, we map the QWZ model onto the type-II hyperbolic lattice which can be described by a tight-binding model (TBM) Hamiltonian:

$$H_{\text{CI}} = \sum_{\langle i,j \rangle} c_i^\dagger J_{ij} c_j + \sum_i (M + 2t_2) c_i^\dagger \sigma_z c_i \tag{1}$$

where $c_i^\dagger$ and $c_i$ are the creation and annihilation operators of electrons at the site $i$, $\sigma_{x,y,z}$ are the Pauli matrices. Matrix $J_{ij} = -[it_1(\sigma_x \cos\theta_{ij} + \sigma_y \sin\theta_{ij}) + t_2\sigma_z]/2$ depicts the hopping between two neighboring connected sites, where $\theta_{ij}$ represents the polar angle of the vector from site $i$ to site $j$ (see label in Fig. 2a), serving as a hopping phase to break the time-reversal symmetry of the system. Note that although the mirror symmetry of the sublattices outside and inside $r_h$ is broken by the current hopping phase setting, the amplitudes of all couplings are identical, ensuring the nature of negative constant curvature. We henceforth set system parameters $t_1 = t_2 = 1$, $M = -1$. We directly diagonalize the Hamiltonian $H_{\text{CI}}$ to obtain the energy spectrum, as shown in Fig. 2c. The bulk energy gap (orange region), indicated by the vanished bulk density of states (DOS) (see Supplementary Note 4) shown in Fig. 2b, is occupied by paired inner (red dots) and outer (blue dots) CESs, whose intensity distributions are displayed in Fig. 2e.

To further verify the nontrivial topological properties of these CESs, we first calculate their corresponding real-space Bott index $C_B$, as shown in Fig. 2d. In the gap region, $C_B = -1$ indicates the nontrivial band topology of the type-II HCI phase (see Supplementary Note 5). Next, to characterize the nonreciprocal and robust transport of these CESs, we inject wave packets $\psi_0(t) = \exp\left(-\frac{t^2}{2\sigma^2}\right)\sin(2\pi\varepsilon_0 t)$ with $\varepsilon_0 = 0.0064$ and $\sigma = 50$ (green stars in Fig. 2f) into the outer and inner edges to investigate their dynamic time evolution and introduce several on-site potential defects (white circles in Fig. 2f) to test their topological protection. Figure 2f presents the instantaneous intensity distributions of $|\psi_0(t)|$ at different times $t = 60, 172, 292, 424$, respectively. It can be seen that the wave packet is well-confined to the outer (inner) edge of the type-II HCI and propagates nonreciprocally along the counterclockwise (CCW) [clockwise (CW)] direction, smoothly passes through the defects without any backscattering. These results prove that both the inner and outer edges of type-II HCIs can support robust and nonreciprocal CESs.

**Dynamic transfer of CESs in modulated type-II HCIs**

The key to achieving the dynamic transfer between the outer and inner CESs lies in introducing interaction and controlling the evolution path of the coupled states in real space or parameter space. In real space, a straightforward paradigm is to completely couple a CCW-propagating CES at the outer edge, through the bulk, to a CW-propagating CES at the inner edge, and vice versa. This implies that these two CESs coalesce as one state with zero power flow, which is the characteristic signature of an APT EP [42-44]. We start by considering a two-level model and selecting the two modes in Fig. 2d as its orthonormal basis $\{|\psi_o\rangle, |\psi_i\rangle\}$, representing the CESs localized at the outer and inner edges, respectively. The CCW and CW power flows of the bases are defined as $\beta_o = |\langle\psi_o|\psi_o\rangle|^2 - |\langle\psi_i|\psi_o\rangle|^2 = 1 = \Delta$ and $\beta_i = |\langle\psi_o|\psi_i\rangle|^2 - |\langle\psi_i|\psi_i\rangle|^2 = -1 = -\Delta$, respectively. To manipulate the interactions between these two counterpropagating CESs, we introduce a radial coupling channel (pink region) consisting of seven modulated couplings (pink short lines) $T_{ij}(\rho) = \rho \cdot J_{ij}$, as illustrated in Fig. 3a. Figure 3b shows the eigenenergy of the system as a function of the modulation strength $\rho$. Remarkably, as $\rho$ increases, the approximately degenerate states $|\psi_{\rho,k}\rangle$ ($k = 1, 2$) gradually split, and their power flows, given by $\beta_k = |\langle\psi_o|\psi_{\rho,k}\rangle|^2 - |\langle\psi_i|\psi_{\rho,k}\rangle|^2$, change from 1 (blue color) or -1 (red color) to 0 (green color) simultaneously. This phenomenon of power flow evolution can be described by a two-level effective Hamiltonian $H_{eff} = \begin{pmatrix} \beta_o & \kappa_{oi} \\ \kappa_{io} & \beta_i \end{pmatrix}$, where the coupling factors are derived as purely imaginary numbers $\kappa_{oi} = \kappa_{io} = i\kappa$ due to the power-flow-difference conservation in the contra-directional coupling process [42-44] (see Supplementary Note 6). Hence, we arrive at the Hamiltonian written as:

$$H_{APT} = \begin{pmatrix} \Delta & i\kappa \\ i\kappa & -\Delta \end{pmatrix} \quad (3)$$

which is APT-S, i.e., $\{H_{APT}, \hat{P}\hat{T}\} = 0$ and gives eigenvalues:

$$\beta = \pm\sqrt{\Delta^2 - \kappa^2} \quad (4)$$

As shown by solid lines in Fig. 3c, in the weakly coupled regime $|\kappa| < |\Delta|$ (pink region), $\beta$ appear as two real numbers, corresponding to an anti-parity-time-symmetry-broken (APT-B) phase which hosts two propagating eigenstates with CCW (blue line) or CW (red line) power flows. Especially, when the coupling reaches a critical point $|\kappa| = |\Delta|$ (yellow dashed line), $\beta$ coalesces to zero, corresponding to an EP (yellow star) and non-propagating eigenstates with vanishing power flow. Additionally, $\beta$ in Fig. 3b can also be retrieved as a function of effective $\kappa$, as demonstrated by circles in Fig. 3b, which exhibits a remarkable quantitative agreement with the analytic results from Eq. (4). Figure 3d shows the EP eigenstates corresponding to the yellow stars in Figs. 3b-3c, which exhibit nearly identical intensity distributions that conform to the coalescence of eigenvectors at the EP, and the equal intensity distribution between the outer and inner edges coincide with the vanishing power flow at the EP.

We then examine the dependence of outer edge to inner edge transmissions on the modulation strength $\rho$. As illustrated in the inset of Fig. 3e, we inject the formerly used wave-packet $\psi_0(t)$ from an outer edge source (green star marked 1), and then evaluate the transmission $S_{21}$ ($S_{31}$) from an outer (inner) edge probe [blue dot marked 2 (red dot marked 3)] for different $\rho$. Figure 3e reveals that, for sufficiently weak $\rho$ ($< 8$), $S_{21}$ ($S_{31}$) remains stable at 1 (0), indicating the wave-packet only propagates along the outer edge. As $\rho$ increases, the system gradually approaches the EP, $S_{21}$ ($S_{31}$) exponentially decays (increases) to 0 (1), signifying the dynamic transfer from the outer CESs to the inner CESs. Figure 3f presents the instantaneous intensity distributions of $|\psi_0(t)|$ at different times t = 40, 84, 104, 292 with $\rho = 101$, respectively. It can be seen that the wave packet initially propagates CCW along the outer edge, and then gradually crosses the coupling channel and continues to propagate CW along the inner edge, thereby achieving the dynamic transfer of CESs in type-II HCIs via the APT phase transition which has never been reported before.

Next, to overcome the challenge of the CESs failing to dynamically transfer under sufficiently small $\rho$ (i.e., far from EP), we harness a new and robust mechanism regardless of modulation strength, namely the dynamic state pumping along an adiabatically evolved path in parameter space, such as LZ single-band pumping. As illustrated in Fig. 4a, we introduce additional non-reciprocal phases $\Phi$ into the couplings $T_{ij}(\rho)$ as a new degree of freedom. For $\rho = 6$, which is an extremely weak modulation strength that cannot induce CES transfer using the APT phase transition. Figure 4b shows the eigenenergy of the coupled outer and inner CESs near zero energy as a function of the non-reciprocal phases $\Phi$, where the blue and red colors represent the weighting of the corresponding eigenstates on the bases $|\psi_o\rangle$ and $|\psi_i\rangle$ with a CES gap (green region) of a size of $\Delta E = 0.000446$ at $\Phi = \pi$. These two gapped bands can be modeled as a two-level LZ Hamiltonian [36,37]:

$$H_{LZ}(\delta\Phi) = \begin{pmatrix} -\alpha\delta\Phi & \Gamma \\ \Gamma & \alpha\delta\Phi \end{pmatrix} \quad (5)$$

where $\Gamma = \Delta E/2$ is determined by the gap size and $\alpha = 0.002$ is a fitting parameter. If the initial state is $|\psi_i\rangle$ at $\Phi = \pi - 1$ (red stars in Figs. 4b-4c), there are two possible routes of state evolution, as illustrated by the dashed red and blue arrows in Fig. 4b. When it follows the red arrow path, the $|\psi_i\rangle$ component always dominates during this process, and the final state (red rhombus in Figs. 4b-4c) remains localized at the inner edge. In contrast, when it is pumped along the blue arrow path, the final state (blue rhombus in Figs. 4b-4c) will be dominated by $|\psi_o\rangle$, leading to the transition from the inner edge to the outer edge and the final states will be localized at the outer edge.

The final state $|\psi_f\rangle$ is a superposition of bases $|\psi_o\rangle$ and $|\psi_i\rangle$ given by $|\psi_f\rangle = O(t)|\psi_o\rangle + I(t)|\psi_i\rangle$, where the composition $O(t)$ and $I(t)$ satisfy:

$$-i\frac{d}{dt}\begin{pmatrix} O(t) \\ I(t) \end{pmatrix} = \begin{pmatrix} \eta t & \Gamma \\ \Gamma & -\eta t \end{pmatrix} \begin{pmatrix} O(t) \\ I(t) \end{pmatrix} \quad (6)$$

where $\eta = \alpha(\Delta\Phi/T)$ characterizes the adiabaticity that depends on the evolution rate $\Delta\Phi/T$, i.e., the ratio of phase range $\Delta\Phi = 2$ and the number of evolution steps $T$. The state evolution starts with the initial state dominated by $|\psi_i\rangle$, i.e., the initial condition is $(0,1)^T$. Using Eq. (6), the weightings of the final state on the bases $|\psi_i\rangle$ and $|\psi_o\rangle$ can be derived as a function of $T$: $P_{|\psi_i\rangle} = I(T)^2 = e^{-\pi\Gamma^2/\eta}$ and $P_{|\psi_o\rangle} = O(T)^2 = 1 - e^{-\pi\Gamma^2/\eta}$ [36,37]. We plot the weightings as solid lines in Fig. 4d, it can be seen that $P_{|\psi_i\rangle}$ and $P_{|\psi_o\rangle}$ intersects at $T_t = \frac{\alpha\Delta\Phi \ln 2}{\pi\Gamma^2} = 17747$ (black dashed line), which is the LZ transition point. When $T \gg T_t$ (blue dashed line in Fig. 4d), the state evolves sufficiently slowly to remain adiabatic and stays on the same band with the final state dominated by $|\psi_o\rangle$, corresponding to the LZ single-band pumping (blue arrow in Fig. 4b). However, when the same process occurs with $T \ll T_t$ (red dashed line in Fig. 4d), the state evolution will be fast enough to be nonadiabatic and tunnel across the energy gap, populating the upper band with the final state dominated by $|\psi_i\rangle$ and corresponding to the LZ tunneling (red arrow in Fig. 4b).

For the initial state (red star) shown in Figs. 4b-4c, we calculate the evolved final state at different evolution rates and determine its weightings on the bases $|\psi_i\rangle$ and $|\psi_o\rangle$ as functions of $T$, as plotted by colored dots in Fig. 4d, which coincide with the results of the LZ model. Figures 4e and 4f display the intensity distributions of the state during the nonadiabatic ($T = T_1 = 1001$) and adiabatic ($T = T_2 = 125001$) evolution, respectively. As predicted in our theoretical analysis, for the LZ tunneling with nonadiabatic evolution (Fig. 4e), the initial inner CESs remain localized at the inner edge, while for the LZ single-band pumping with adiabatic evolution (Fig. 4f), the initial inner CESs pump across the bulk to the outer edge, achieving the dynamic transfer of the CESs in type-II HCIs. Given that the characteristic radius $r_h$ plays a pivotal role in type-II hyperbolic lattices, we have discussed the influence of $r_h$ on the dynamic transfer of CESs via two mechanisms. We find that the transfer efficiency of

CESs in the process of APT phase transition is not significantly affected by $r_h$, whereas in the process of LZ single-band pumping, a larger $r_h$ leads to a lower transfer efficiency (see Supplementary Note 7).

**Conclusion**

In conclusion, we have theoretically demonstrated the nontrivial CESs and their dynamic transfer in topological type-II hyperbolic lattices. By extending the QWZ model to the type-II hyperbolic space, we achieve a type-II HCI featuring inner and outer CESs with opposite chirality, in contrast to its type-I counterpart which only has outer CESs with a single chirality. More interestingly, we demonstrate that the counterpropagating CESs at the outer and inner edges of type-II HCIs can be dynamically transferred via two distinct mechanisms: the completely reversed power-flow conversion at an APT-S EP and the adiabatic evolution along a single-band in LZ model. Our work thus significantly expands the research scope of hyperbolic topological physics and establishes a paradigm for exploring novel topological effects in more complex non-Euclidean spaces. We envision the dynamic transfer of hyperbolic surface states or hyperbolic hinge states that can be achieved in three-dimensional stacked type-II HCIs. Given current experimental techniques, the dynamic transfer of the CESs in topological type-II hyperbolic lattices can be readily realized in circuit quantum electrodynamics [14], electric circuits [15,17,21,22], integrated nanophotonic chips [23], and nonreciprocal networks [24]. Our work may inspire potential applications in robust and compact hyperbolic topological devices by leveraging dynamic manipulations of the hyperbolic topological states, such as hyperbolic topological laser [45,46] and optical frequency combs [47-49].

**Data availability**

All data that support the plots within this paper and other findings of this work are available at https://zenodo.org/records/14747644.

**Code availability**

All the codes used to generate and/or analyze the data in this work are available at https://zenodo.org/records/14747644.

**Reference**

[1] Hasan, M. Z. & Kane, C. L. Colloquium: Topological insulators. *Rev. Mod. Phys.* **82**, 3045–3067 (2010).
[2] Qi, X.-L. & Zhang, S.-C. Topological insulators and superconductors. *Rev. Mod. Phys.* **83**, 1057–1110 (2011).
[3] Lu, L., Joannopoulos, J. D. & Soljačić, M. Topological photonics. *Nat. Photonics* **8**, 821–829 (2014).
[4] Ozawa, T. et al. Topological photonics. *Rev. Mod. Phys.* **91**, 015006 (2019).
[5] Yang, Z., Gao, F., Shi, X., Lin, X., Gao, Z., Chong, Y. & Zhang, B. Topological acoustics. *Phys. Rev. Lett.* **114**, 114301 (2015).

[6] Xue, H., Yang, Y. & Zhang, B. Topological acoustics. *Nat. Rev. Mater.* **7**, 974–990 (2022).
[7] Ma, G., Xiao, M. & Chan, C. T. Topological phases in acoustic and mechanical systems. *Nat. Rev. Phys.* **1**, 281–294 (2019).
[8] Süsstrunk, R. & Huber, S. D. Observation of phononic helical edge states in a mechanical topological insulator. *Science* **349**, 47–50 (2015).
[9] Ningyuan, J., Owens, C., Sommer, A., Schuster, D. & Simon, J. Time- and site-resolved dynamics in a topological circuit. *Phys. Rev. X.* **5**, 021031 (2015).
[10] Imhof, S. et al. Topolectrical-circuit realization of topological corner modes. *Nat. Phys.* **14**, 925–929 (2018).
[11] Lee, C. H. et al. Topolectrical circuits. *Commun. Phys.* **1**, 39 (2018).
[12] Zhang, Z. et al. Diffusion metamaterials. *Rev. Mod. Phys.* **5**, 218-235 (2023).
[13] Xu, G. et al. Diffusive topological transport in spatiotemporal thermal lattices. *Nat. Phys.* **18**, 450–456 (2022).
[14] Kollár, A. J., Fitzpatrick, M. & Houck, A. A. Hyperbolic lattices in circuit quantum electrodynamics. *Nature* **571**, 45 (2019).
[15] Lenggenhager, P. M. et al. Simulating hyperbolic space on a circuit board. *Nat. Commun.* **13**, 4373 (2022).
[16] Yu, S., Piao, X. & Park, N. Topological hyperbolic lattices. *Phys. Rev. Lett.* **125**, 053901 306 (2020).
[17] Zhang, W., Yuan, H., Sun, N., Sun, H. & Zhang, X. Observation of novel topological states in hyperbolic lattices. *Nat. Commun.* **13**, 2937 (2022).
[18] Urwyler, D. M. et al. Hyperbolic topological band insulators. *Phys. Rev. Lett.* **129**, 246402 (2022).
[19] Liu, Z.-R., Hua, C.-B., Peng, T. & Zhou, B. Chern insulator in a hyperbolic lattice. *Phys. Rev. B* **105**, 245301 (2022).
[20] Stegmaier, A., Upreti, L. K., Thomale, R. & Boettcher, I. Universality of Hofstadter butterflies on hyperbolic lattices. *Phys. Rev. Lett.* **128**, 166402 (2022).
[21] Zhang, W., Di, F., Zheng, X., Sun, H. & Zhang, X. Hyperbolic band topology with non-trivial second Chern numbers. *Nat. Commun.* **14**, 1083 (2023).
[22] Chen, A. et al. Hyperbolic matter in electrical circuits with tunable complex phases. *Nat. Commun.* **14**, 622 (2022).
[23] Huang, L. et al. Hyperbolic photonic topological insulators. *Nat. Commun.* **15**, 1647 (2024).
[24] Chen, Q. et al. Anomalous and Chern topological waves in hyperbolic networks. *Nat. Commun.* **15**, 2293 (2024).
[25] Tummuru, T. et al. Hyperbolic non-Abelian semimetal. *Phys. Rev. Lett.* **132**, 206601 (2024).
[26] Tao, Y.-L. & Xu, Y. Higher-order topological hyperbolic lattices. *Phys. Rev. B* **107**, 184201 (2023).
[27] Liu, Z.-R., Hua, C.-B., Peng, T., Chen, R. & Zhou, B. Higher-order topological insulators in hyperbolic lattices. *Phys. Rev. B* **107**, 125302 (2023).
[28] Thouless, D. J. Quantization of particle transport. *Phys. Rev. B* **27**, 6083-6087 (1983).

[29] Citro, R. & Aidelsburger, M. Thouless pumping and topology. *Nat. Rev. Phys.* **5**, 87-101 (2023).
[30] Laughlin, R. B. Quantized Hall conductivity in two dimensions. *Phys. Rev. B* **23**, 5632 (1981).
[31] Fabre, A. et al. Laughlin's topological charge pump in an atomic Hall cylinder. *Phys. Rev. Lett.* **128**, 173202 (2022).
[32] Kawamura, M. et al. Laughlin charge pumping in a quantum anomalous Hall insulator. *Nat. Phys.* **19**, 333-337 (2023).
[33] Suh, J. et al. Photonic topological spin pump in synthetic frequency dimensions. *Phys. Rev. Lett.* **132**, 033803 (2024).
[34] Landau, L. D. & Lifshitz, E. M. Quantum mechanics: non-relativistic theory (Elsevier Science, New York, 1991).
[35] Zener, C. Non-adiabatic crossing of energy levels. *Proc. R. Soc. Lond. A* **137**, 696–702 (1932).
[36] Chen, Z. G. et al. Landau-Zener transition in the dynamic transfer of acoustic topological states. *Phys. Rev. Lett.* **126**, 054301 (2021).
[37] Xu, B. C. et al. Topological Landau–Zener nanophotonic circuits. *Adv. Photonics* **5**, **3**, 036005 (2023).
[38] Lohse, M. et al. Exploring 4D quantum Hall physics with a 2D topological charge pump. *Nature* **553**, 55-58 (2018).
[39] Ziberberg, O. et al. Photonic topological boundary pumping as a probe of 4D quantum Hall physics. *Nature* **553**, 59-62 (2018).
[40] Chen, J. et al. AdS/CFT correspondence in hyperbolic lattices. arXiv: 2305.04862.
[41] Dey, S. et al. Simulating holographic conformal field theories on hyperbolic lattices. *Phys. Rev. Lett.* **133**, 061603 (2024).
[42] Yariv, A. Coupled-mode theory for guided-wave optics. *IEEE J. Quantum Electron.* **9**, 919 (1973).
[43] Choi, Y. et al. Observation of an anti-PT-symmetric exceptional point and energy-difference conserving dynamics in electrical circuit resonators. *Nat. Commun.* **9**, 2182 (2018).
[44] Xie, X. R. et al. Harnessing anti-parity-time phase transition in coupled topological photonic valley waveguides. *Adv. Funct. Mater.* **33**, 2302197 (2023).
[45] Harari, G. et al. Topological insulator laser: Theory, Science 359, eaar4003 (2018).
[46] Bandres, M. A. et al, Topological insulator laser: Experiment, Science 359, eaar4005 (2018).
[47] Mittal, S. et al. Topological frequency combs and nested temporal solitons, Nat. Phys. 17, 1169-1176 (2021).
[48] Flower, C. J. et al. Observation of topological frequency combs, Science 384, 1356-1361 (2024).
[49] Huang, L. et al. Hyperbolic Topological Frequency Combs, ACS Photonics 4c01546 (2024).

**Acknowledgments**

Z.G. acknowledges the finding from the National Natural Science Foundation of China under grants No. 62375118, 62361166627, and 12104211, Guangdong Basic and Applied Basic Research Foundation under grant No.2024A1515012770, Shenzhen Science and Technology Innovation Commission under grants No. 202208151111105001 and 202308073000209, High level of special funds under grant No. G03034K004.

**Authors Contributions**

Z.G. initiated the project. J.M.C. and L.Y.Y. performed numerical analysis. J.M.C., L.Y.Y., and Z.G. analyzed data. J.M.C. and Z.G. wrote and revised the manuscript. Z.G. supervised the project.

**Competing Interests**

The authors declare no competing interests.

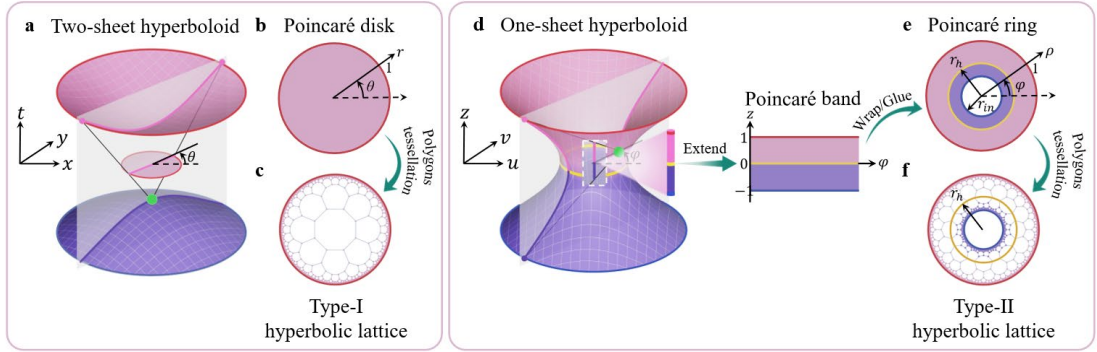

**Fig. 1 | Type-I and type-II hyperbolic lattices. a** One brunch of a two-sheet hyperboloid ($t^2 - x^2 - y^2 = 1$) in (2+1)-dimensional ($x, y, t$) Minkowski spacetime is mapped onto a Poincaré disk ($z = 0$) by stereographic projection through the point $(0, 0, -1)$ (green dot). **b** Poincaré disk with unit radius. **c** A type-I hyperbolic lattice with $\{8, 3\}$-tessellation constructed by discretizing the Poincaré disk with octagons. **d** One-sheet hyperboloid ($u^2 + v^2 - z^2 = 1$) in (1+2)-dimensional ($u, v, z$) Minkowski spacetime is mapped onto a set of overlapped line segments at $z$ axis by stereographic projection through the point $(0,1,0)$ (green dot). Inset displays a Poincaré band configuration formed by extending the line segments along angular position $\varphi$. **e** Poincaré ring with unit outer radius, inner radius $r_{in}$, and characteristic radius $r_h$ generated by wrapping and gluing the Poincaré band in **d**. **f** A type-II hyperbolic lattice with $\{r_h, 8, 3\}$-tessellation constructed by discretizing the Poincaré ring with octagons.

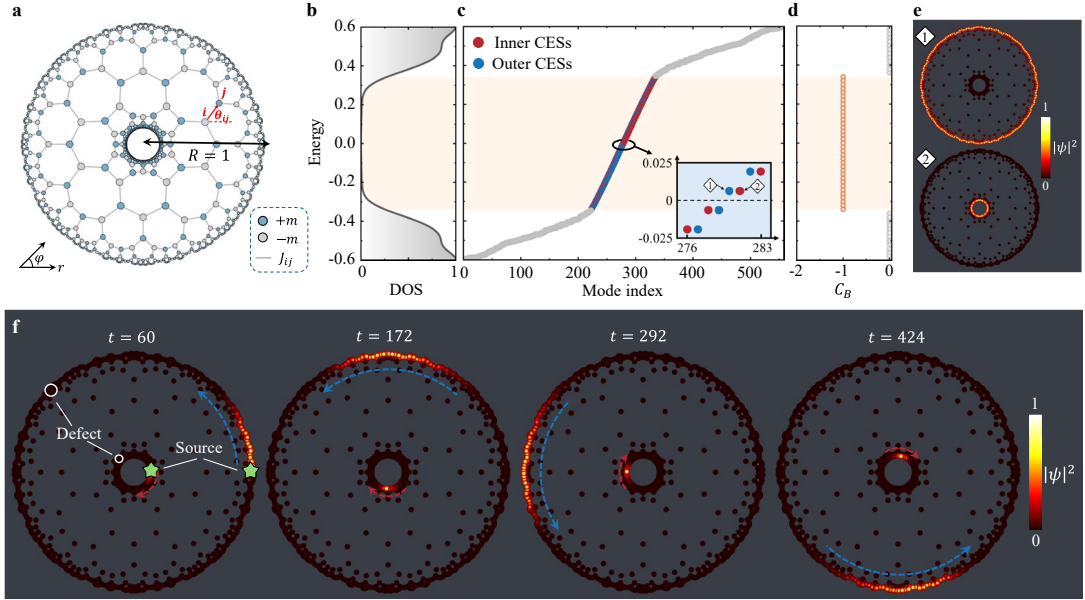

**Fig. 2 | Type-II hyperbolic Chern insulator (HCI). a** Schematic illustration of Qi-Wu-Zhang model on a type-II hyperbolic {0.365, 8, 3} lattice. **b** Bulk density of state (DOS) of the Type-II HCI in **a**, the energy region with vanished DOS indicates the bulk energy gap (orange region). **c** Energy spectrum of the Type-II HCI in **a**, inner (red dots) and outer (blue dots) chiral edge states (CESs) fill the energy gap. The inset displays the zoomed-in view of the CESs near zero energy. **d** Real-space Bott index $C_B$ as a function of energy. **e** Intensity distributions of the outer and inner CESs marked in the inset of **c**. **f** Instantaneous intensity distributions of wave-packet excited by two edge sources (green stars) at time 60, 172, 292, and 424, respectively.

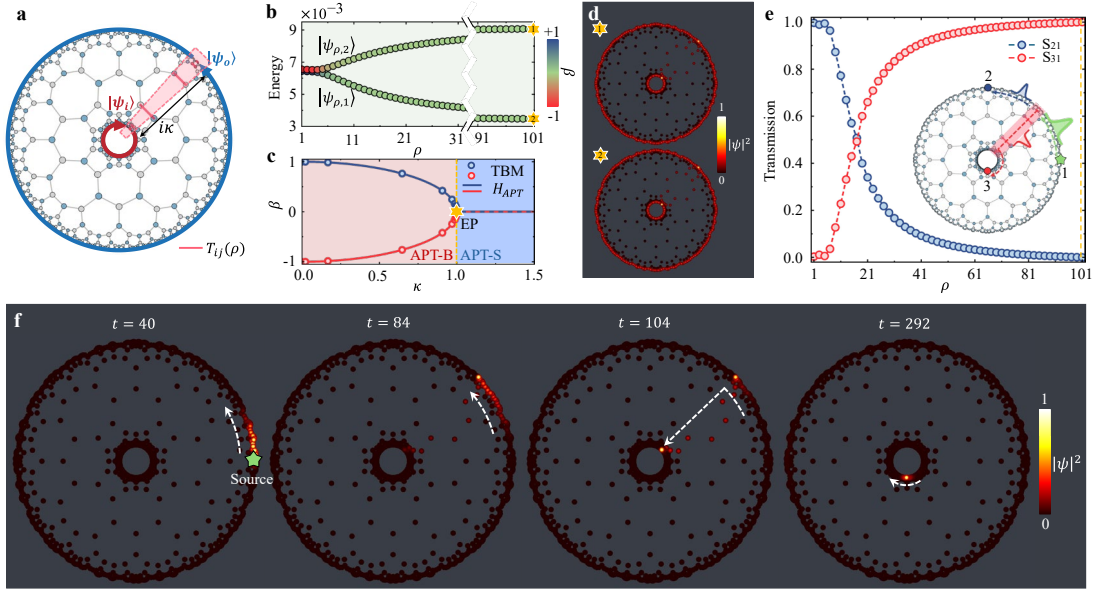

**Fig. 3 | Dynamic transfer of chiral edge states via anti-parity-time (APT) phase transition.** **a** Schematic illustration of a modulated type-II hyperbolic Chern insulator for implementing an effective two-level anti-parity-time systematic (APT-S) model. **b** Eigenenergy of the system in **a** as a function of the modulation strength $\rho$, where the color indicates the power flow $\beta$ of each eigenstate. **c** $\beta$ as a function of $\kappa$. The solid lines (dots) are analytic (numerical) results calculated from the APT Hamiltonian $H_{APT}$ (tight-binding model calculations). Pink and blue regions denote the anti-parity-time-symmetry-broken (APT-B) and APT-S phase, respectively. The yellow star represents an APT exceptional point (EP). **d** Intensity distributions of quasi-EP eigenstates marked by yellow stars in **b** and **c**. **e** Wave packet transmissions of outer-outer ($S_{21}$) and outer-inner ($S_{31}$) edge transports as a function of $\rho$. Inset illustrates that a wave packet excited by an outer edge source (green star), is spited into two wave packets by the channel, which are detected by outer (blue dot) and inner (red dot) edge probes. **f** Instantaneous intensity distributions of wave-packet at time 40, 84, 104, and 292, respectively, for $\rho = 100$ marked by yellow dashed line in **e**.

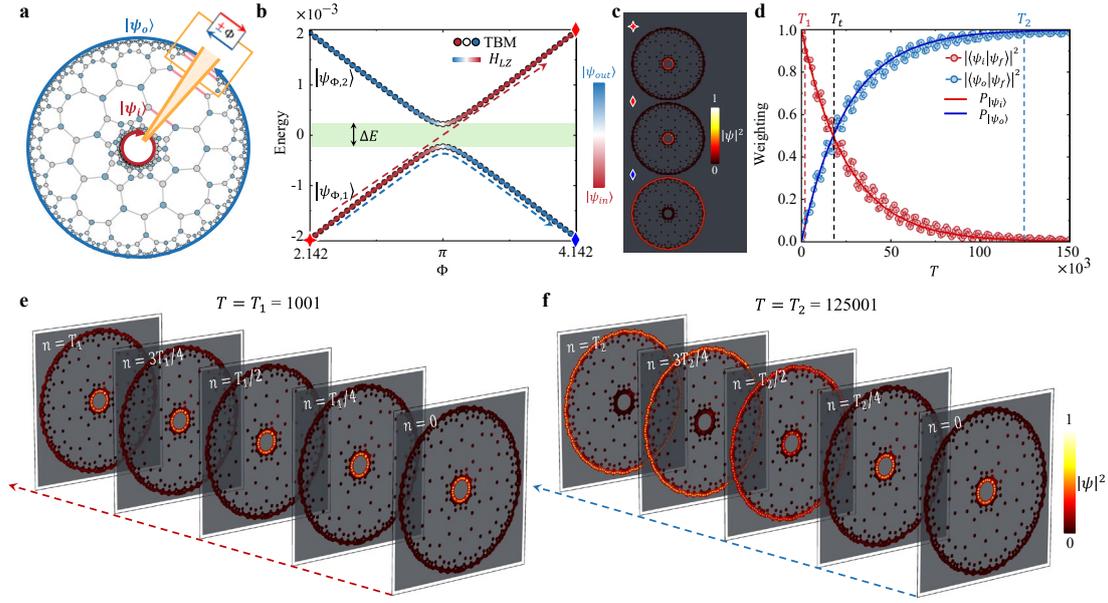

**Fig. 4 | Dynamic transfer of chiral edge states via Landau-Zener (LZ) single-band pumping**. **a** Schematic illustration of a modulated type-II hyperbolic Chern insulator for implementing an effective two-level LZ model. **b** Eigenenergy of the system in **a** as a function of modulation phase $\Phi$, where the color indicates the weightings of each eigenstate on the bases $|\psi_i\rangle$ (red color) and $|\psi_o\rangle$ (blue color). The solid lines (dots) are analytic (numerical) results calculated from the LZ Hamiltonian $H_{LZ}$ (tight-binding model calculations). The green region represents the energy gap. **c** The intensity distributions of the initial state (red star) and the two possible final states (red and blue rhombuses) along different state evolution paths. **d** The weightings of final state $|\psi_f\rangle$ on the bases $|\psi_i\rangle$ and $|\psi_o\rangle$ as a function of the number of evolution step $T$. **e** and **f** The changes in intensity distribution of state during a nonadiabatic (**e**) and an adiabatic (**f**) evolution, corresponding to the red and blue dashed lines in **d**, respectively.